\documentclass[12pt]{article}

\usepackage{mathptmx}
\usepackage{amssymb,amsmath,amsbsy,amsfonts}
\usepackage{graphicx}
\usepackage[left=1cm,right=1cm,top=2cm,bottom=2cm]{geometry}
\usepackage{braket}
\usepackage{color}
\usepackage{mathrsfs}
\usepackage{cite}
\usepackage{hyperref}
\usepackage{physics}
\usepackage{subfig}
\usepackage{float}
\usepackage{hyperref}
\hypersetup{colorlinks = true,
	linkcolor = red,
	anchorcolor = blue,
	citecolor = blue,
	filecolor = blue,
	urlcolor = blue}
	
\DeclareMathOperator\arctanh{arctanh}

\begin{document}

\title{Exact solution for the time dependent non-Hermitian generalized Swanson oscillator.}

\author{B.M. Villegas-Martínez$^*$, H.M. Moya-Cessa and F. Soto-Eguibar\\
	\small Instituto Nacional de Astrofísica, Óptica y Electrónica, INAOE \\
	\small {Calle Luis Enrique Erro 1, Santa María Tonantzintla, Puebla, 72840 Mexico}\\
	$^*$\small {Corresponding author: bvillegas@inaoep.mx}}

\date{\today}

\maketitle

\begin{abstract}
We produce an exact solution of the Schrödinger equation for the generalized time dependent Swanson oscillator. The system studied is a non-Hermitian setup characterized by time dependent complex coefficients. The exact solution is obtained by  applying two  transformations and under the right choice of the relevant parameters. Consequently, the model is reduced to a time independent harmonic oscillator.
\end{abstract}

\section{Introduction}
$\mathcal{PT}$-symmetric non-Hermitian systems have gained a flurry of interest due to its inherent potential of exhibiting a real spectrum, notwithstanding the non-hermiticity of the Hamiltonians. Historically, it was Bender’s \cite{1,2,3,4} pioneering work who marked the grounds for these systems, with lacking the Hermiticity condition, to acquire a new meaning with the constraint of $\mathcal{PT}$-symmetry, i.e., the discrete space-time symmetries of parity ($\mathcal{P}$) and time-reversal ($\mathcal{T}$). Since then, many prototypical examples of the above symmetry-based scenario have been presented in the literature\cite{5}, ranging from theoretical scrutiny in open quantum systems\cite{6} or quantum optics\cite{7} to powerful applications in optics such as perfect laser absorbers \cite{8}, spatial optical switches\cite{9}, among others. Nonetheless, only a few special classes of quantum $\mathcal{PT}$-models are fully exactly solvable; an outstanding example that fulfills this requirement is the popular non-Hermitian quadratic Hamiltonian proposed by Swanson\cite{10}; the properties of the Swanson oscillator have been extensively undertaken in \cite{10}. Subsequently, studies related to supersymmetric realization of such models and \textit{q}-deformation boson algebras have been carried out in \cite{11,12,13,14,15}. Quite remarkably, several authors have found that the time independent Swanson model can be mapped to a harmonic oscillator by performing a gauge-like transformation, Bogoliubov transformation or non-unitary transformation \cite{16,17}. Exact solutions have been found for the time dependent case by using Lewis and Riesenfeld time dependent invariants and time-dependent non-unitary transformation \cite{17.1,17.2}. Among recent works, Zelaya {\it et al.} \cite{18} have presented an extension of the Swanson model with arbitrary time dependent real parameters; there, the Schrödinger equation for a special case of this model produces a  generalization of the Caldirola-Kanai oscillator. In particular, in the concrete generalization of the Swanson oscillator provided in Reference \cite{18}, it is missing a study of the non-Hermitian case defined by arbitrary time dependent {\it complex-valued} functions. Therefore, influenced by Zelaya’s {\it et al.} seminal landmark article, we proceed in this work to expand the above analysis to study time dependent non-unitary transformations that allow  to obtain exact solutions of the generalized Swanson oscillator. Our purpose is not to engage in a comprehensive review; rather, we want to show that under an adequate choice of the complex time dependent functions, it is possible to obtain an exact solution of the Hamiltonian; we present the model and the needed transformations for both scenarios where the Hamiltonian system may be time independent or time dependent.\\
The layout of the paper is as follows: In Section 2, we place emphasis on solving the Schrödinger equation of the time independent generalized Swanson oscillator with real coefficients instead of complex ones. We show that the system can be rendered to the Non-hermitian forced harmonic oscillator by a suitable choice of  the coefficients, whose exact solution can be easily obtained. Section 3 is devoted to the scenario where the Hamiltonian is time dependent possessing complex coefficient functions; under the suitable selection of these functions, we present the exact solution of two specific cases concerning to the Non-Hermitian Caldirola–Kanai and the generalized Swanson system with time-dependent growing mass. Finally, conclusions and discussion of the work are presented in Section 4.

\section{Time-independent generalized Swanson oscillator}
The starting point is the non-Hermitian time-independent Hamiltonian of the generalized Swanson oscillator defined by \cite{18,18.1}
\begin{equation} \label{1a}
\frac{1}{\omega_{0}}\hat{H}_{GSW}=\theta \left(\hat{a}^{\dagger} \hat{a} + \hat{a} \hat{a}^{\dagger}\right) + \alpha_{1} \hat{a}^{\dagger} + \beta_{1} \hat{a} + \alpha_{2} \hat{a}^{\dagger2} + \beta_{2} \hat{a}^2,
\end{equation}
where $\hat{a}$ and $\hat{a}^{\dagger}$ are the bosonic annihilation and creation operators of the standard harmonic oscillator, with real time-independent functions $\theta$ and $\alpha_{j} \neq \beta_{j}$ with $j=1,2$, being $\omega_{0}>0$ a constant with units of frequency. The Hamiltonian $\hat{H}_{GSW}$ can be recast in terms of position and momentum operators by the well-known relationships $\hat{a}=\sqrt{\frac{m_{0} \omega_{0}}{2 }}\left(\hat{x}+i\frac{\hat{p}}{m_{0} \omega_{0}}\right)$ and $\hat{a}^{\dagger}=\sqrt{\frac{m_{0} \omega_{0}}{2 }}\left(\hat{x}-i\frac{\hat{p}}{m_{0} \omega_{0}}\right)$ with constant mass $m_{0}$; one can easily prove that
\begin{equation} \label{2a}
\hat{H}_{GSW}=\nu_{1}\hat{p}^2 + \nu_{2} \hat{x}^2 + i \nu_{3}\left(\hat{x}\hat{p}+\hat{p}\hat{x}\right) + i \nu_{4} \hat{p} + \nu_{5} \hat{x},
\end{equation}
where
\begin{align} \label{3a}
\nu_{1}=&\frac{2\theta -\left(\alpha_{2}+\beta_{2}\right)}{2 m_{0}}, \quad \nu_{2}=\frac{m_{0}\omega^2_{0}}{2}\left(2\theta + \alpha_{2} + \beta_{2}\right),  \quad \nu_{3}=\frac{\omega_{0}}{2}\left( \beta_{2}-\alpha_{2}\right), 
\nonumber \\ 
\nu_{4}=& \sqrt{\frac{\omega_{0}}{2 m_{0}}} \left(\beta_{1}-\alpha_{1}\right), \quad \qquad \nu_{5}=\sqrt{\frac{m_{0}\omega^3_{0}}{2}}\left(\alpha_{1}+\beta_{1}\right).
\end{align}
It is important to mention that the linear terms of $\hat{x}$ and $\hat{p}$ make this kind of physical system non-$\mathcal{PT}$-symmetric; this can be verified with the usual form of parity and time reversal operators, $\mathcal{P}$: $\hat{x}\rightarrow -\hat{x}$; $\hat{p}\rightarrow -\hat{p}$, $\mathcal{T}:\hat{x}\rightarrow \hat{x}$; $\hat{p}\rightarrow -\hat{p}; i \rightarrow -i$. For the case $\nu_{4}=\nu_{5}=0$, we return back to the original $\mathcal{PT}$-symmetric Swanson oscillator.

\subsection{Non-hermitian forced harmonic oscillator}
Before proceeding to solve exactly the Hamiltonian \eqref{2a}, we begin by discussing the particularly simple case when  $\alpha_{2}=\beta_{2}=0$, and which is commonly called the non-Hermitian and non-$\mathcal{PT}$-symmetric forced harmonic oscillator; this system is described by the Schrödinger equation
\begin{equation} \label{4a}
i\frac{d}{dt}\ket{\psi(t)}=\left[\frac{\theta}{m_{0}}\left(\hat{p}^2 + m^2_{0} \omega^2_{0}  \hat{x}^2 \right)  + i \nu_{4} \hat{p} + \nu_{5} \hat{x}\right]\ket{\psi(t)}.
\end{equation}
If we introduce the non-unitary transformation $\ket{\psi(t)}=\hat{\eta} \ket{\phi(t)}$ with $\hat{\eta}=\exp\big\lbrace -\frac{\left[m_{0}\omega_{0} \left(\alpha_{1}-\beta_{1}\right) \hat{x}-i\left(\alpha_{1}+\beta_{1}\right)\hat{p}\right]}{2 \theta \sqrt{2 m_{0}\omega_{0}}}  \big\rbrace$ and use the formula $e^{\hat{A}}\hat{B}e^{-\hat{A}}=\hat{B}+ \left[\hat{A},\hat{B}\right]+ \frac{1}{2!}\left[\hat{A}, \left[\hat{A},\hat{B}\right]\right]+\frac{1}{3!}\left[\hat{A}, \left[\hat{A}, \left[\hat{A},\hat{B}\right]\right]\right]+ \ldots$ \cite{18.2,18.3}, we get the new evolution equation
\begin{equation} \label{5a}
i\frac{d}{dt}\ket{\phi(t)}=\left[\frac{\theta}{m_{0}} \left(\hat{p}^2 +m^2_{0} \omega^2_{0} \hat{x}^2 \right)-\frac{\omega_{0}\alpha_{1}\beta_{1}}{2\theta}\right]\ket{\phi(t)},
\end{equation}
which is nothing else than a harmonic oscillator displaced by the quantity $\frac{\omega_{0}\alpha_{1}\beta_{1}}{2\theta}$ and with an energy $E_{n}=\theta \omega_{0} \left(2n+1\right)-\frac{\omega_{0}\alpha_{1}\beta_{1}}{2\theta}$. Integrating the resulting expression over $t$ and transforming back to the original representation $\ket{\psi(t)}$, one finds the exact solution
\begin{equation} \label{6a}
\ket{\psi(t)}= \exp\left(i\frac{\omega_{0}\alpha_{1}\beta_{1}}{2\theta} t \right) \hat{\eta} \exp\left[-i\frac{\theta}{m_{0}} \left(\hat{p}^2 +m^2_{0} \omega^2_{0} \hat{x}^2 \right)t \right] \hat{\eta}^{-1} \ket{\psi(0)}.
\end{equation}

\subsection{Exact solution for the generalized Swanson oscillator}
In this Section, we are interested in finding the solution of the Schrödinger equation associated with the Hamiltonian $\hat{H}_{GSW}$, expression \eqref{2a}; the exact solution can be obtained by performing two non-unitary transformations. First, we apply $\ket{\psi(t)}=\hat{\eta}_{1} \ket{\phi(t)}$, where
\begin{equation} \label{7a}
\hat{\eta}_{1}=\exp\Bigg\lbrace i \kappa \left[\sqrt{\frac{m_{0}\omega_{0}}{2}} \sin\left(\vartheta\right) \hat{x} +\frac{\cos\left(\vartheta\right)}{\sqrt{2m_{0}\omega_{0}}} \hat{p} \right]\Bigg\rbrace,    
\end{equation}
with
\begin{equation} \label{8a}
\kappa=\sqrt{\frac{m_{0} \omega_{0}}{2}} \; \frac{\nu_{3} \nu_{4} + \nu_{1} \nu_{5}}{\nu_{1} \nu_{2} + \nu_{3}^2 }\sqrt{1-\frac{\left(\nu_{2} \nu_{4} - \nu_{3} \nu_{5}\right)^2}{m^2_{0} \omega^2_{0}\left(\nu_{3} \nu_{4} + \nu_{1} \nu_{5}\right)^2}},
\qquad 
\vartheta=i \arctanh\left[\frac{ \frac{\nu_{3} \nu_{5}}{\nu_{4}}-\nu_{2}}{m_{0}\omega_{0} \left(\nu_{3} + \frac{\nu_{1}\nu_{5}}{\nu_{4}}\right)} \right],
\end{equation}
and we get the transformed Schrödinger equation
\begin{equation} \label{9a}
i \frac{d}{dt}\ket{\phi(t)}=\left[\nu_{1} \hat{p}^2 + \nu_{2} \hat{x}^2 + i \nu_{3} \left(\hat{x}\hat{p}+\hat{p}\hat{x}\right) + \frac{\nu_{2} \nu^2_{4}-\nu_{5}\left(\nu_{1} \nu_{5} + 2 \nu_{3}\nu_{4} \right)}{4\left(\nu_{1}\nu_{2} + \nu^2_{3}\right)} \right]\ket{\phi(t)}.
\end{equation}
The transformation induced by \eqref{7a} gets rid of the linear terms in $\hat{x}$ and $\hat{p}$; in the absence of these terms, the Hamiltonian acquires the form of the $\mathcal{PT}$-symmetric Swanson oscillator. Secondly, we perform the transformation $\ket{\phi(t)}=\hat{\eta}_{2}\ket{\chi(t)}$, which is similar to the reported by \cite{19}, where
\begin{align} \label{10a}
\hat{\eta}_{2}=&\exp  \left\lbrace \ln\left[\frac{\sqrt{\nu_{1}\nu_{2}+\nu^2_{3}}}{\left(\frac{\nu_{2}}{2 m_{0}\omega_{0}} + \frac{\nu_{1}m_{0}\omega_{0}}{2}\right) + \sqrt{\left(\frac{\nu_{2}}{2 m_{0}\omega_{0}} - \frac{\nu_{1}m_{0}\omega_{0}}{2}\right)^2-\nu^2_{3}}}\right] 
\right. \nonumber \\ & \left. 
\times \left[\frac{\nu_{3}\left(\frac{m_{0}\omega_{0}}{2} \hat{x}^2 -\frac{\hat{p}^2}{2 m_{0}\omega_{0}}\right) + i\left(\frac{\nu_{2}}{2 m_{0}\omega_{0}} +\frac{\nu_{1}m_{0}\omega_{0}}{2}\right) \left(\hat{x} \hat{p} + \hat{p}\hat{x}\right)}{2\sqrt{\left(\frac{\nu_{2}}{2 m_{0}\omega_{0}} - \frac{\nu_{1}m_{0}\omega_{0}}{2}\right)^2-\nu^2_{3}}}\right] 
\right\rbrace,
\end{align}
and Eq.~\eqref{9a} is converted into
\begin{equation}  \label{11a}
i\frac{d}{dt}\ket{\chi(t)}=\hat{\tilde{H}} \ket{\chi(t)} =\left[ \frac{\tilde{\omega}}{2 m_{0}\omega_{0}} \left(\hat{p}^2 + m^2_{0}\omega^2_{0} \hat{x}^2 \right)  + \delta \right]\ket{\chi(t)};
\end{equation}
hence, the transformation $\eta_{2}$ lead us to the Schrödinger equation of a harmonic oscillator with 
\begin{equation} \label{12a}
\tilde{\omega}=2\sqrt{\nu_{1}\nu_{2}+\nu^2_{3}}, \qquad  \delta= \frac{\nu_{2} \nu^2_{4}-\nu_{5}\left(\nu_{1} \nu_{5} + 2 \nu_{3}\nu_{4} \right)}{4\left(\nu_{1}\nu_{2} + \nu^2_{3}\right)}.
\end{equation}
One, therefore, obtains the energy spectrum of the original Hamiltonian $\hat{H}_{GSW}$ from the eigenequation of the Hamiltonian $\hat{\tilde{H}}$, i.e, from $\hat{\tilde{H}}\ket{n}=E_{n}\ket{n}$, where
\begin{equation} \label{13a}
E_{n}=\sqrt{\nu_{1}\nu_{2}+\nu^2_{3}} \left(2n+1\right)  + \frac{\nu_{2} \nu^2_{4}-\nu_{5}\left(\nu_{1} \nu_{5} + 2 \nu_{3}\nu_{4} \right)}{4\left(\nu_{1}\nu_{2} + \nu^2_{3}\right)},
\qquad n=0,1,2,....
\end{equation}
The eigenfunctions of $\hat{H}_{GSW}$ can be derived from the eigenfunctions of the harmonic oscillator via the association
\begin{equation} \label{14a}
    \hat{\tilde{H}}=\hat{\eta}^{-1}_{2} \hat{\eta}^{-1}_{1}\hat{H}_{GSW}\hat{\eta}_{1}\eta_{2}\ket{n}=E_{n}\ket{n} \quad \Leftrightarrow \quad \hat{H}_{GSW}\ket{\tilde{n}}=E_{n}\ket{\tilde{n}}
\end{equation}
where $\ket{\tilde{n}}=\hat{\eta}_{1}\hat{\eta}_{2}\ket{n}$. The above results successfully check the correctness of the results derived in \cite{20}. \\
Finally, the exact solution of the Schrödinger equation corresponding to $\hat{H}_{GSW}$ can be written as
\begin{equation} \label{15a}
\ket{\psi(t)}=\exp\left(-i\delta t\right)\hat{\eta}_{1} \hat{\eta}_{2} \exp\left[-\frac{i\tilde{\omega}t}{2 m_{0}\omega_{0}} \left(\hat{p}^2 + m^2_{0}\omega^2_{0} \hat{x}^2 \right) \right] \hat{\eta}^{-1}_{1} \hat{\eta}^{-1}_{2}\ket{\psi(0)}.
\end{equation}

\section{Generalized Swanson Hamiltonian with time-dependent complex coefficients}
In what follows, we focus on the time-dependent generalized Swanson Hamiltonian \cite{18,18.4,21}
\begin{equation} \label{16a}
\frac{1}{\omega_{0}}\hat{H}_{GSW}(t)=\theta(t) \left(\hat{a}^{\dagger} \hat{a} + \hat{a} \hat{a}^{\dagger}\right) + \alpha_{1}(t) \hat{a}^{\dagger} + \beta_{1}(t) \hat{a} + \alpha_{2}(t) \hat{a}^{\dagger2} + \beta_{2}(t) \hat{a}^2 + V_{0}(t),
\end{equation}
with complex time-dependent functions $\theta(t)$, $V_{0}(t)$, $\alpha_{j} (t)$ and $\beta_{j}(t)$ with $j=1,2$. Following the same procedure as in Section 2, and using the same notation reported in \cite{18}, we rewrite the Hamiltonian \eqref{16a} in terms of momentum and position operators as
\begin{equation} \label{17a}
\hat{H}_{GSW}(t)=\frac{\hat{p}^2}{2m(t)} + \frac{m(t) \omega^2(t)}{2}\hat{x}^2 + i \frac{\Omega(t)}{2} \left(\hat{x} \hat{p} + \hat{p} \hat{x}\right) + i \nu(t) \hat{p} + F(t) \hat{x} + \omega_{0} V_{0}(t),
\end{equation}
where the new time-dependent $m(t)$, $\omega^2(t)$, $\Omega(t)$, $\nu(t)$ and $F(t)$ functions are defined by
\begin{align} \label{18a}
m(t)&= \frac{m_{0} }{2 \theta(t)-\left[\alpha_{2}(t) +\beta_{2}(t)\right]},  &\omega^2(t)&= \omega^2_{0} \lbrace 4 \theta^2 (t) -\left[\alpha_{2}(t) +\beta_{2}(t)\right]^2\rbrace, 
\nonumber \\
\Omega(t)&=-\omega_{0} \left[\alpha_{2}(t)-\beta_{2}(t)\right],  &\nu(t)&=-\sqrt{\frac{ \omega_{0}}{2 m_{0}}} \left[\alpha_{1}(t) -\beta_{1}(t)\right], 
\nonumber \\
F(t)&=\sqrt{\frac{m_{0}\omega^3_{0}}{2}} \left[\alpha_{1}(t) +\beta_{1}(t)\right]. 
\end{align}

\subsection{Non-Hermitian Caldirola–Kanai: Case $\alpha_{2}=\beta_{2}$}
Clearly the time-dependent configuration of $\hat{H}_{GSW}(t)$ is non-Hermitian when $\alpha_{j} (t) \neq \beta^{*}_{j}(t)$ and also it is not $\mathcal{PT}$-symmetric. Since the solution of the generalized Swanson oscillator is completely determined by the choice of above time-dependent functions; then, it is interesting to consider some special cases for which the time-dependent Schr\"{o}dinger equation associated to $\hat{H}_{GSW}(t)$ has exact closed form solutions. Let us consider a model generated by the time-dependent functions 
\begin{align} \label{19a}
\theta(t)&= \frac{1}{2} \cos\left(2 \varGamma t\right), & \alpha_{1}(t)&= \sqrt{\frac{2 m_{0}}{\omega_{0}}} \nu_{0}\cos\left(\varGamma t\right), \nonumber \\
\beta_{1}(t)&= i \sqrt{\frac{2 m_{0}}{\omega_{0}}} \nu_{0} \sin\left(\varGamma t\right), & \alpha_{2}(t)&=\beta_{2}(t)=\frac{i}{2} \sin\left(2 \varGamma t\right);
\end{align}
this lead us to the non-Hermitian Caldirola–Kanai Hamiltonian system
\begin{equation} \label{20a}
\hat{H}_{NCK}(t)=\frac{\hat{p}^2}{2 m_{0}} \exp\left(-2 i \varGamma t\right) + \frac{m_{0}\omega^2_{0}}{2} \hat{x}^2 \exp\left(2 i \varGamma t\right) + i \nu_{0} \hat{p} \exp\left(-i \varGamma t\right) + \nu_{0} m_{0} \omega_{0} \exp\left(i \varGamma t\right) \hat{x},
\end{equation}
whose mass depends exponentially on time, i.e., $m(t)= m_{0} \exp\left(2 i \varGamma t\right)$ \cite{21.1,21.2}. For the sake of simplicity, we have chosen $V_{0}(t)=0$. \\
In order to solve the corresponding Schrödinger equation of the above Hamiltonian, we consider the non-unitary transformation $\ket{\psi(t)}=\exp\left[-\frac{ \varGamma t}{2} \left(\hat{x}\hat{p}+ \hat{p}\hat{x}\right)\right]\ket{\phi(t)}$. The idea behind the transformation is to remove the temporal dependence of Hamiltonian \eqref{20a}, to get
\begin{equation} \label{21a}
i\frac{d}{dt}\ket{\phi(t)}=\left[ \frac{\hat{p}^2}{2 m_{0}} + \frac{m_{0}\omega^2_{0}}{2} \hat{x}^2 + i \frac{ \varGamma}{2} \left(\hat{x} \hat{p} + \hat{p} \hat{x}\right) + i\nu_{0} \hat{p} + \nu_{0} m_{0} \omega_{0} \hat{x}  \right]\ket{\phi(t)}.
\end{equation}
Strikingly, by identifying $\frac{1}{2m_{0}}\rightarrow \nu_{1}$, $\frac{m_{0}\omega^2_{0}}{2}\rightarrow \nu_{2}$, $\frac{\varGamma}{2} \rightarrow \nu_{3}$, $\nu_{0} \rightarrow \nu_{4}$, $\nu_{0} m_{0} \omega_{0} \rightarrow \nu_{5}$ and according to the procedure outlined in Subsection 2.2, we can write the exact solution as
\begin{equation} \label{22a}
\ket{\psi(t)}=\exp\left(i\frac{\varGamma \nu^2_{0} m_{0} \omega_{0}}{\omega^2_{0} + \varGamma^2} t\right)\hat{\eta}_{1} \hat{\eta}_{2} \exp\left[-i t\frac{ \sqrt{\omega^2_{0} + \varGamma^2}}{2 m_{0}\omega_{0}} \left(\hat{p}^2 + m^2_{0}\omega^2_{0} \hat{x}^2 \right) \right] \hat{\eta}^{-1}_{1} \hat{\eta}^{-1}_{2}\ket{\psi(0)},
\end{equation}
with the energy spectrum given by
\begin{equation} \label{23a}
E_{n}=\frac{\sqrt{\omega^2_{0} + \varGamma^2}}{2}\left(2n+1\right) - \frac{ \varGamma \nu^2_{0} m_{0}\omega_{0}}{\omega^2_{0} + \varGamma^2}.
\end{equation}

\subsection{Generalized Swanson Hamiltonian with complex mass growing with time: Case $\alpha_{2}\neq \beta_{2}$}
\begin{figure} [H]
\subfloat(a)
{\includegraphics[width=0.45\textwidth]{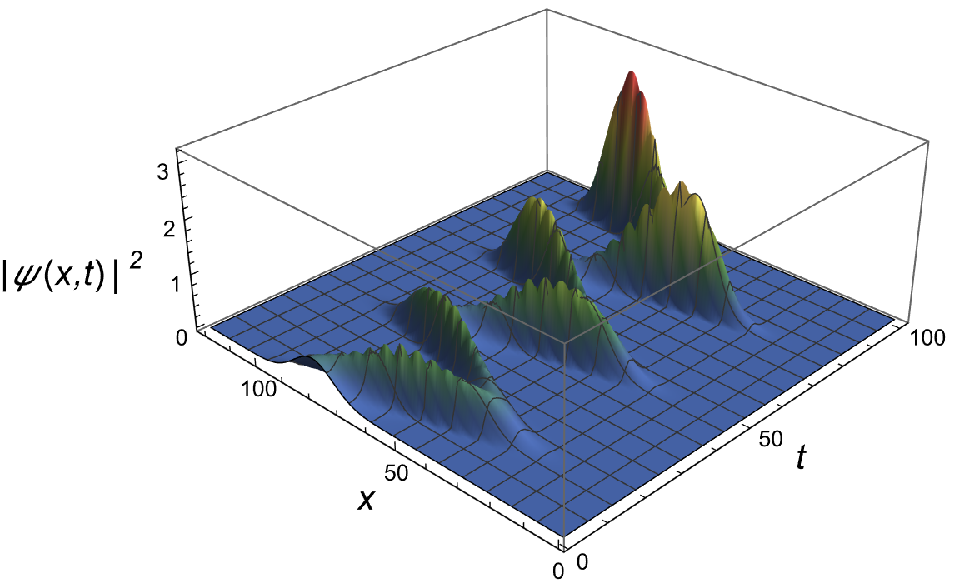}}\hfill
\subfloat(b)
{\includegraphics[width=0.45\textwidth]{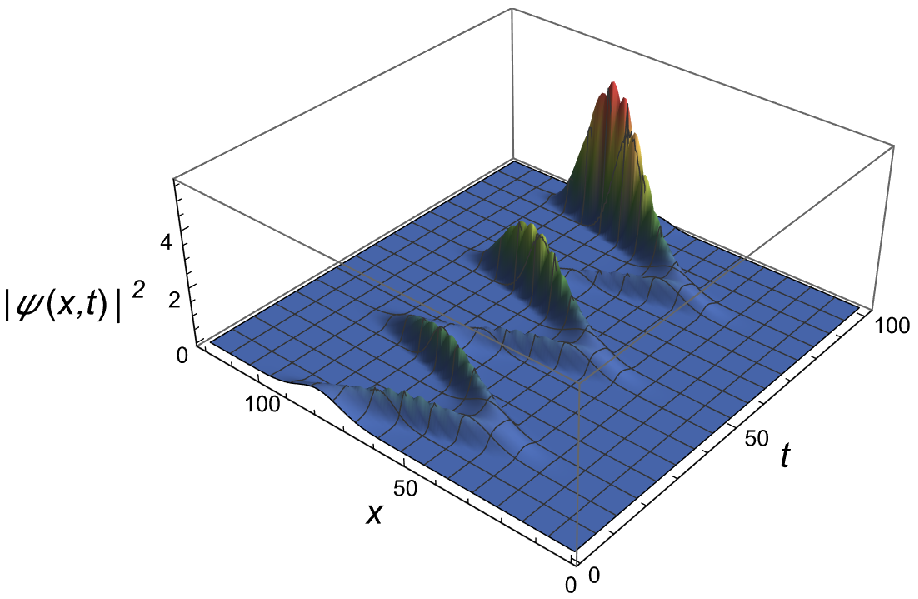}} 
\caption{Amplitude probability distribution $\abs{\psi(x,t)}^2$. The left graph (a) is plotted using the parameters $m_{0}=1$, $\omega_{0}=2$, $\sigma=1/\sqrt{2}$, $x_{0}=2$, $\Omega_{0}=0.015$, $\nu_{0}=0.00015$. For (b) case, the set of values are $m_{0}=1$, $\omega_{0}=2$, $\sigma=1/\sqrt{2}$, $x_{0}=2$, $\Omega_{0}=0.01$ and $\nu_{0}=0.015$.}
\label{f1}
\end{figure}

In the following, we define the complex time-dependent functions
\begin{align} \label{24a}
\theta(t)&=\frac{1 + \left(1-2 i t \Omega_{0}\right)^2}{4\left(1-2 i t \Omega_{0}\right)}, 
&\alpha_{1}(t)&= -\frac{\nu_{0}\sqrt{\frac{m_{0}}{2\omega_{0}}}\left(1+ i t \omega_{0}+2 t^2 \omega_{0} \Omega_{0}\right)}{\sqrt{1-2 i t \Omega_{0}}}, 
\nonumber \\
\beta_{1}(t)&=\frac{\nu_{0}\sqrt{\frac{m_{0}}{2\omega_{0}}}\left(1 -i t \omega_{0}- 2t^2 \omega_{0}\Omega_{0}\right)}{\sqrt{1-2 i t \Omega_{0}}}, 
&\alpha_{2}(t)&= -\frac{\Omega_{0}\left(1+ 2 i t \omega_{0} + 2 t^2 \omega_{0} \Omega_{0}\right)}{2 \omega_{0}\left(1-2 i t \Omega_{0}\right)}, 
\nonumber\\
\beta_{2}(t)&= \frac{\Omega_{0}\left(1- 2 i t \omega_{0} - 2 t^2 \omega_{0} \Omega_{0}\right)}{2 \omega_{0}\left(1-2 i t \Omega_{0}\right)}, & V_{0}(t)&=- \frac{1}{2}\gamma^2(t)m_{0}\omega^2_{0} \left(1-2 i t \Omega_{0}\right),
\end{align}
with $\Omega_{0}\geq0$ and $\nu_{0}\geq 0$. Then the Hamiltonian \eqref{17a} acquires the form
\begin{align} \label{25a}
\hat{H}_{GSW}(t)=&\frac{\hat{p}^2}{2 m_{0}\left(1-2 i t \Omega_{0}\right)} + \frac{m_{0} \omega^2_{0}\left(1-2 i t \Omega_{0}\right) }{2}\hat{x}^2 + i \frac{\Omega_{0}}{2\left(1-2 i t \Omega_{0}\right)} \left(\hat{x} \hat{p} + \hat{p} \hat{x}\right) + i \frac{\nu_{0}}{\sqrt{1-2 i t \Omega_{0}}} \hat{p} \nonumber \\
& -it m_{0} \nu_{0} \omega^2_{0} \sqrt{1-2 i t \Omega_{0}} \hat{x} - \frac{1}{2}\gamma^2(t)m_{0}\omega^2_{0} \left(1-2 i t \Omega_{0}\right).
\end{align}
It is not difficult to infer that the set of complex functions yields to an oscillator with complex mass which increases with time; nevertheless, one can perform multiple configurations of the time dependent functions to analyse different mass choices. In order to solve the Schrödinger equation associated with $\hat{H}_{GSW}(t)$, $i\frac{d}{dt} \ket{\psi(t)}=\hat{H}_{SW}(t)\ket{\psi(t)}$, we do two non-unitary transformations. First, consider $\ket{\psi(t)}=\exp\left[\gamma(t) \hat{p}\right]\ket{\phi(t)}$ with $\gamma(t)=\frac{\nu_{0} t}{\sqrt{1-2 i t \Omega_{0}}}$; the main goal of this transformation is to remove the linear terms in $\hat{x}$ and $\hat{p}$, similarly to the time-independent case of Subsection 2.2; the transformed Schrödinger equation becomes
\begin{align} \label{26a}
i\frac{d}{dt}\ket{\phi(t)}=&\left[ \frac{\hat{p}^2}{2 m_{0}\left(1-2 i t \Omega_{0}\right)} + \frac{m_{0} \omega^2_{0}\left(1-2 i t \Omega_{0}\right) }{2} \hat{x}^2  +  i \frac{\Omega_{0}}{2\left(1-2 i t \Omega_{0}\right)} \left(\hat{x} \hat{p} + \hat{p} \hat{x}\right)
\right]\ket{\phi(t)};
\end{align}
note that the non-Hermitian time-dependent Hamiltonian inside of brackets is $\mathcal{PT}$-symmetric, since it is invariant under the combined operations of space-time inversion: $\mathcal{P}:\hat{x}\rightarrow -\hat{x}$; $\mathcal{T}:t\rightarrow -t, i\rightarrow -i$\cite{21.3,21.4,21.5}.
Secondly, we want to eliminate the time-dependence and get rid of the squeeze term $\left(\hat{x} \hat{p} + \hat{p} \hat{x}\right)$; to do this, we consider a second and final transformation $\ket{\phi(t)}=\exp\left[ \frac{\delta(t)}{2} \left(\hat{x} \hat{p}+\hat{p}\hat{x}\right)\right] \ket{\chi(t)}$ with $\delta(t)=\frac{i}{2}\ln\left(1-2it\Omega_{0}\right)$, to get
\begin{align} \label{27a}
i \frac{d}{dt}\ket{\chi(t)}= \left( \frac{\hat{p}^2}{2 m_{0}} + \frac{m_{0}\omega^2_{0}}{2} \hat{x}^2 \right) \ket{\chi(t)}.
\end{align}
Evidently, our second transformation maps the time-dependent generalized harmonic oscillator to an ordinary time-independent harmonic oscillator.\\
Finally, the exact solution is obtained by using the inverse transformations previously established on this Subsection
\begin{equation} \label{28a}
\ket{\psi(t)}=\exp\left(\frac{\nu_{0} t}{\sqrt{1-2 i t \Omega_{0}}} \hat{p}\right)\exp\left[ \frac{i}{4}\ln\left(1-2it\Omega_{0}\right) \left(\hat{x} \hat{p}+\hat{p}\hat{x}\right)\right] \exp\left[-\frac{it}{2m_{0}}\left(\hat{p}^2 + m^2_{0}\omega^2_{0} \hat{x}^2\right)\right] \ket{\psi(0)}.
\end{equation}
It is worth mentioning that the exact solution presented above is composed by the product of three exponential operators, being the first one (from left to right) a displacement-like operator followed by a squeezed-like similarity operator which acts over the propagator related to the harmonic oscillator; this factorization order also takes place in the time-independence case in Eq.~\eqref{15a}. Moreover, one can use results from Ref.\cite{22} to factorize the exponential operator of the harmonic oscillator and obtain the following form
\begin{align} \label{29a}
\ket{\psi(t)}=&\left(1-2 i t \Omega_{0}\right)^{1/4} \exp\left(\frac{\nu_{0} t}{\sqrt{1-2 i t \Omega_{0}}} \hat{p}\right) \exp\left[\frac{i}{2}\ln\left(1-2 i t \Omega_{0}\right) \hat{x} \hat{p}\right] \exp\left[-i\frac{\tan\left(\omega_{0} t/2\right)}{2 m_{0} \omega_{0}} \hat{p}^2\right] \nonumber\\
& \times \exp\left[-i \frac{m_{0} \omega_{0}}{2} \sin\left(\omega_{0}t\right) \hat{x}^2\right]  \exp\left[-i\frac{\tan\left(\omega_{0} t/2\right)}{2 m_{0} \omega_{0}} \hat{p}^2\right] \ket{\psi(0)},
\end{align}
whose solution in the coordinate representation, $\psi(x,t)=\bra{x}\ket{\psi(t)}$, reads as
\begin{align} \label{30a}
\psi(x,t)=&\left(1-2 i t \Omega_{0}\right)^{1/4} \exp\left(-i\frac{\nu_{0} t}{\sqrt{1-2 i t \Omega_{0}}} \frac{d}{dx} \right) \exp\left[\frac{1}{2}\ln\left(1-2 i t \Omega_{0}\right) x \frac{d}{dx}\right] \exp\left[i\frac{\tan\left(\omega_{0} t/2\right)}{2 m_{0} \omega_{0}} \frac{d^2}{dx^2}\right] \nonumber\\
& \times \exp\left[-i \frac{m_{0} \omega_{0}}{2} \sin\left(\omega_{0}t\right) x^2\right]  \exp\left[i\frac{\tan\left(\omega_{0} t/2\right)}{2 m_{0} \omega_{0}} \frac{d^2}{dx^2}\right] \psi(x,0).
\end{align}
The next step is to evaluate the exact solution with an initial condition. In this case we choose as an initial wave function $\psi(x,0)=\frac{1}{\left(\sigma \sqrt{2\pi}\right)^{1/2}} \exp \left[{-\frac{\left(x-x_{0}\right)^2}{4\sigma^2}}\right]$ which it is displaced a distance $x_{0}$ from the origin, being $\sigma$ the width of the wavepacket. Then, the product of exponential operators from the Harmonic oscillator acting on initial state leads to
\begin{align} \label{31a}
\psi(x,t)=&\frac{\left(1-2 i t \Omega_{0}\right)^{1/4} \exp\left(-i\frac{\nu_{0} t}{\sqrt{1-2 i t \Omega_{0}}} \frac{d}{dx} \right) }{\left(\sigma \sqrt{2\pi}\right)^{1/2} \sqrt{\cos\left(\omega_{0}t\right) + \left(2i/m_{0}\omega_{0}\right)\left(1/4\sigma^2\right)\sin\left(\omega_{0}t\right)}}  \exp\left[\frac{1}{2}\ln\left(1-2 i t \Omega_{0}\right) x \frac{d}{dx}\right] \nonumber\\
& \times \exp\left[ \frac{\left(x_{0}/4\sigma^2\right)^2}{\left(i m_{0} \omega_{0}/2\right)\tan\left(\omega_{0}t/2\right)+\left(1/4\sigma^2\right)} -\left(x^2_{0}/4\sigma^2\right) \right] \exp\left[-i\frac{m_{0}\omega_{0}}{2} \tan\left(\omega_{0}t/2\right) x^2\right]  \nonumber\\
& \times \exp \bigg\lbrace -\frac{ \left(i m_0 \omega_0/2\right)\tan\left(\omega_{0}t/2\right) + \left(1/4\sigma^2\right)}{\cos\left(\omega_{0}t\right) + \left(2i/m_{0}\omega_{0}\right)\left(1/4\sigma^2\right)\sin\left(\omega_{0}t\right)} \left[x-\frac{x_{0}}{1+ 2i\sigma^2 m_{0}\omega_{0} \tan\left(\omega_{0}t/2\right)}\right]^2 \bigg\rbrace,
\end{align}
and using the definition of the dilatation operator, $\exp\left(\lambda x\frac{d}{dx}\right)f\left(x\right)=f\left(e^{\lambda} x \right)$, and of the translation operator, $\exp\left(\lambda \frac{d}{dx}\right) f\left(x\right)=f\left(x+\lambda\right)$, acting on a given function $f(x)$ \cite{23}, we arrive to
\begin{align} \label{32a}
\psi(x,t)&=\frac{\left(1-2 i t \Omega_{0}\right)^{1/4} \exp\left[ \frac{\left(x_{0}/4\sigma^2\right)^2}{\left(i m_{0} \omega_{0}/2\right)\tan\left(\omega_{0}t/2\right)+\left(1/4\sigma^2\right)} -\left(x^2_{0}/4\sigma^2\right) \right] }{\left(\sigma \sqrt{2\pi}\right)^{1/2} \sqrt{\cos\left(\omega_{0}t\right) + \left(2i/m_{0}\omega_{0}\right)\left(1/4\sigma^2\right)\sin\left(\omega_{0}t\right)}} \nonumber\\
& \times  \exp\left[-i\frac{m_{0}\omega_{0}}{2} \tan\left(\omega_{0}t/2\right)\left(1-2i t \Omega_{0}\right) \left(x-i\frac{\nu_{0}t}{\sqrt{1-2it\Omega_{0}}}\right)^2\right]  \nonumber\\
& \times \exp \bigg\lbrace -\frac{ \left(i m_0 \omega_0/2\right)\tan\left(\omega_{0}t/2\right) + \left(1/4\sigma^2\right)}{\cos\left(\omega_{0}t\right) + \left(2i/m_{0}\omega_{0}\right)\left(1/4\sigma^2\right)\sin\left(\omega_{0}t\right)} \left[x\sqrt{1-2it\Omega_{0}}-i \nu_{0}t-\frac{x_{0}}{1+ 2i\sigma^2 m_{0}\omega_{0} \tan\left(\omega_{0}t/2\right)}\right]^2 \bigg\rbrace.
\end{align}
This wavefuntion is a solution of the Schrödinger equation $i\frac{d}{dt}\ket{\psi(t)}=\hat{H}_{GSW}(t)\ket{\psi(t)}$. In Fig.\ref{f1}, we have plotted the probability density $\abs{\psi(x,t)}^2$; the results depicted there correspond to $m_{0}=1$, $\omega_{0}=2$, $\sigma=1/\sqrt{2}$ and $x_{0}=2$ with two choices of $\Omega_{0}$ and $\nu_{0}$, which are $\Omega_{0}=0.015$, $\nu_{0}=0.00015$ and $\Omega_{0}=0.01$ and $\nu_{0}=0.015$. As Fig.\ref{f1} reveals the probability density amplitudes grow without bound as we increase the value of $\Omega_{0}$ and $\nu_{0}$ (see Fig.1 b); essentially, the influence of $\Omega_{0}$ and $\nu_{0}$ over $t$ changes gradually the oscillating behavior of the wave packet as time increases. 

\section{Conclusions}
In summary, we have studied the generalized version of the Swanson Hamiltonian in the time dependent and independent cases. Specifically, in the time-independent scenario, we notice that the spectrum energy and eigenfunctions of the non-Hermitian oscillator Hamiltonian $\hat{H}_{GSW}$ may be mapped to a harmonic oscillator by non-unitary transformations. Meanwhile, in the time dependent case, the degree of difficulty to determine the exact form of the solution of the Schrödinger equation for the Hamiltonian $\hat{H}_{GSW}$ depends on the special elections of the time-dependent functions involved. In particular, the judicious choice of the time-dependent or independent parameters lead to the Hamiltonian $\hat{H}_{GSW}$, which can be converted into relevant sub-classes of non-hermitian Hamiltonians with real spectrum. For instance, in the time-independent case with real coefficients, we have demonstrated that $\hat{H}_{GSW}$ can then be transformed into the non-hermitian forced harmonic oscillator. On the other hand, in the case of complex time-dependent functions, we have reached two models concerning the Non-Hermitian Caldirola–Kanai and the $\hat{H}_{GSW}$ with complex mass growing with time. In any of these cases, we found both systems become exactly solvable by a direct application of time dependent non-unitary transformations that allowed us to avoid the use of time dependent invariants or point transformations.

\section{Acknowledgment}
B.M. Villegas-Martínez wishes to express his gratitude to CONACyT as well as to the National Institute of Astrophysics, Optics and Electronics (INAOE) for financial support.

\end{document}